\documentclass[%
 reprint,
superscriptaddress,
 amsmath,amssymb,
 aps, singlecolumn
]{revtex4-1}

\usepackage{graphicx}% Include figure files
\usepackage{dcolumn}% Align table columns on decimal point
\usepackage{bm}% bold math

\newcommand{\sech}[0]{{\rm sech}}

\begin{document}

\title{Breathing dissipative solitons in optical microresonators}

\author{E.~Lucas}
\thanks{E.L. and M.K. contributed equally to this work}
\affiliation{{\'E}cole Polytechnique F{\'e}d{\'e}rale de Lausanne (EPFL), CH-1015 Lausanne,
Switzerland}

\author{M.~Karpov}
\thanks{E.L. and M.K. contributed equally to this work}
\affiliation{{\'E}cole Polytechnique F{\'e}d{\'e}rale de Lausanne (EPFL), CH-1015 Lausanne,
Switzerland}

\author{H.~Guo}
\affiliation{{\'E}cole Polytechnique F{\'e}d{\'e}rale de Lausanne (EPFL), CH-1015 Lausanne,
Switzerland}

\author{M.L.~Gorodetsky}
\email[]{michael.gorodetsky@gmail.com}
\affiliation{Faculty of Physics, M.V. Lomonosov Moscow State University, 119991
Moscow, Russia}
\affiliation{Russian Quantum Center, Skolkovo 143025, Russia}

\author{T.J.~Kippenberg}
\email[]{tobias.kippenberg@epfl.ch}
\affiliation{{\'E}cole Polytechnique F{\'e}d{\'e}rale de Lausanne (EPFL), CH-1015 Lausanne,
Switzerland}

\date{\today}% It is always \today, today,

\begin{abstract}
Dissipative solitons are self-localized structures resulting from a double balance between dispersion and nonlinearity as well as dissipation and a driving force. They occur in a wide variety of fields ranging from optics, hydrodynamics to chemistry and biology.
Recently, significant interest has focused on their temporal realization in driven optical microresonators, known as dissipative Kerr solitons. They provide access to coherent, chip-scale optical frequency combs, which have already been employed in optical metrology, data communication and spectroscopy. 
Such Kerr resonator systems can exhibit numerous localized intracavity patterns and provide rich insights into nonlinear dynamics.
A particular class of solutions consists of breathing dissipative solitons, representing pulses with oscillating amplitude and duration, for which no comprehensive understanding has been presented to date.
Here, we observe and study single and multiple breathing dissipative solitons in two different microresonator platforms: crystalline $\mathrm{MgF_2}$ resonator and $\mathrm{Si_3N_4}$ integrated microring. We report a deterministic route to access the breathing state, which allowed for a detailed exploration of the breathing dynamics. In particular, we establish the link between the breathing frequency and two system control parameters -- effective pump laser detuning and pump power. Using a fast detection, we present a direct observation of the spatiotemporal dynamics of individual solitons, revealing irregular oscillations and switching. An understanding of breathing solitons is not only of fundamental interest concerning nonlinear systems close to critical transition, but also relevant for applications to prevent breather-induced instabilities in soliton-based frequency combs.
\end{abstract}

\maketitle

\section{Introduction}
Dissipative solitons are localized solutions of the damped driven nonlinear Schr\"odinger equation, occurring in a wide variety of disciplines including plasma physics, matter waves, optics, chemistry and biology \cite{akhmediev2008dissipative}. Recently, the generation of temporal dissipative Kerr solitons (DKS) through parametric conversion in optical microresonators \cite{herr2014soliton} triggered a substantial interest. DKS constitute a way to generate coherent optical frequency combs with large repetition rates in the microwave domain, having a vast application potential. Indeed, they have already been employed in a growing number of proof-of-concept experiments, including coherent terabit telecommunications \cite{Pfeifle20tbs}, coherent receivers \cite{marin2016microresonator}, dual-comb spectroscopy \cite{dualcomb2016vahala} and for the realization of a microwave-to-optical link via self-referencing \cite{Jost2014link,brasch2016self}.
Furthermore, nonlinear microresonators appeared as a suitable platform to study DKS properties and dynamics. Despite the apparent simplicity of the system, microresonators can support a rich variety of field patterns, as reported both in numerical simulations \cite{Barashenkov1996,gelens2014eigenvalue,chembo2014stability,zhou2015stability} and in experiments \cite{xue2015darksol,herr2014soliton,Yi2015soliton,brasch2016photonic}, including bright dissipative Kerr solitons \cite{ herr2014soliton,Yi2015soliton,brasch2016photonic}, dark pulses \cite{xue2015darksol}, platicons \cite{Lobanov2015platicons}, Turing patterns \cite{Coillet2013a}, or soliton crystals \cite{cole2016solitoncrystal}.
Some of these localized patterns can exhibit a rich panel of dynamical instabilities. In particular, bright DKS can undergo \emph{breathing}, i.e. a periodic variation in their duration and amplitude \cite{Matsko2012breathers,Leo2013,bao2016FPU,yu2016breather}.
Breathing dissipative solitons are related to the Fermi-Pasta-Ulam recurrence \cite{bao2016FPU} -- a paradoxical evolution of nonlinearly coupled oscillators, that periodically return to the initial state \cite{fermi1955FPU} -- similar to Kuznetsov-Ma \cite{Kuznetsov1977,Ma1979} and Akhmediev breathers \cite{Akhmediev1987} in conservative nonlinear systems.

Breathing dissipative solitons were first demonstrated in fiber cavities by Leo et al. \cite{Leo2013}.
However, the experimental observation of breathers in optical microresonators has posed a significant challenge, due to the non-trivial soliton generation process \cite{herr2014soliton, brasch2016photonic}, the thermal nonlinearity that may impact the effective laser detuning \cite{carmon2004thermalstability, karpov2016universal}, and high repetition rates ($>10$~GHz) that make direct time-resolved observations difficult. Two groups have recently reported the observation of breathers in microresonators \cite{bao2016FPU,yu2016breather}, where nevertheless, a variety of open questions regarding the breathing regime accessibility and its dynamics remains unexplored. In particular, the oscillation frequency (breathing frequency) has been reported to vary significantly over different platforms, but no clear relation between this key property and the system parameters has emerged to date.
Understanding the breather regime is not only of fundamental interest, but a necessity for applications. The accurate knowledge of the conditions for breather existence allows the prevention of extreme events \cite{coillet2014roguewaves}, ensuring a stable operation of DKS-based microresonator devices, and avoiding excess noise induced by breathing \cite{lucas2016study}.

Here, we present a comprehensive analysis of breathing dissipative solitons in microresonators. First, we demonstrate a deterministic route to access and characterize breathing solitons in two microresonator platforms: crystalline $\rm MgF_2$ whispering gallery mode resonator and $\mathrm{Si_3N_4}$ integrated microresonators. Second, through experimental and theoretical investigations, we perform a detailed exploration of the breathing regime, revealing a link between breathing frequency and the driving laser parameters. Third, we map the breathers' existence range and study its dependence on the pump power. Fourth, we present for the first time to the best of our knowledge, a time-resolved observation of the intracavity pattern evolution in optical microresonators, that reveals the non-stationary breathing dynamics and enables to track the behavior of individual soliton pulses, even with several solitons in the cavity.

\section{Results}

\noindent \textbf{Deterministic access to dissipative breathing solitons.}
The nonlinear dynamics of optical field in continuous wave (CW)-laser-driven microresonators in the presence of the Kerr nonlinearity can be very accurately described using a system of nonlinear coupled mode equations \cite{Chembo2010, Herr2012}, demonstrating almost perfect correspondence with experimental data \cite{lucas2016study}. This system of equations may be considered as a discrete Fourier transform of the damped driven Nonlinear Schr\"odinger equation (NLSE) \cite{herr2014soliton}: 
\begin{eqnarray}
i\frac{\partial\Psi}{\partial\tau}+\frac{1}{2}\frac{\partial^{2}\Psi}{\partial\theta^{2}}+|\Psi|^{2}\Psi=(-i+\zeta_{0})\Psi+if.\label{eq:nls}
\end{eqnarray}
Here $\Psi(\tau,\phi)$ is the normalized intracavity waveform, $\theta$ is the dimensionless longitudinal coordinate,  and $\tau$ the normalized time. Equation \eqref{eq:nls} is usually terme d in optics as Lugiato-Lefever equation (LLE) \cite{Lugiato1987}, where a transverse coordinate is used instead of a longitudinal one in our case.
The nonlinear dynamics of the system is determined by two parameters: the normalized pump power $f$ and detuning $\zeta_0$, defined as \cite{herr2014soliton}:
\begin{align}
    f &= \sqrt{\dfrac{8 g \eta P_{\rm in}}{\kappa^2 \hbar \omega_0}} \ , & \zeta_0 = \dfrac{2\delta\omega}{\kappa} \ ,
\end{align}
where $\kappa$ denotes the loaded resonator linewidth ($Q=\omega_{0}/\kappa$, loaded quality factor), $\eta=\kappa_{\mathrm{ex}}/\kappa$ the coupling coefficient, $P_{\rm in}$ the pump power, $\omega_{0}$ the pumped resonance frequency and $\delta\omega = 2\pi\delta = \omega_{0}-\omega_{p}$ is the detuning of the pump laser from this resonance. The nonlinearity is described via $g=\hbar\omega_{0}^{2}cn_{2}/n_{0}^{2}V_{\mathrm{eff}}$ giving the Kerr frequency shift per photon, with the effective refractive index $n_{0}$, nonlinear refractive index $n_{2}$, and the effective optical mode volume $V_{\mathrm{eff}}$.

A similar equation was first analyzed in plasma physics \cite{Nozaki1984,Nozaki1985}. These early studies demonstrated that stable soliton attractors exist within a certain range of effective detuning in the red-detuned regime ($\zeta_0>0$). It was also shown that, in addition to stable solitons, time-periodic solitons (i.e. breathers) and chaotic states are possible \cite{Barashenkov1996}.
Extensive numerical analysis and charting of the parameter space of \eqref{eq:nls} \cite{Barashenkov1996,Leo2013,gelens2014eigenvalue,chembo2014stability} revealed that the breathing region is located close to the low-detuning boundary of the soliton existence range. Theoretically the transition from stationary solitons to oscillating breathers results from a Hopf bifurcation that arises  above certain pump power level \cite{Leo2013}.

\begin{figure*} 
\includegraphics[width = 2 \columnwidth]{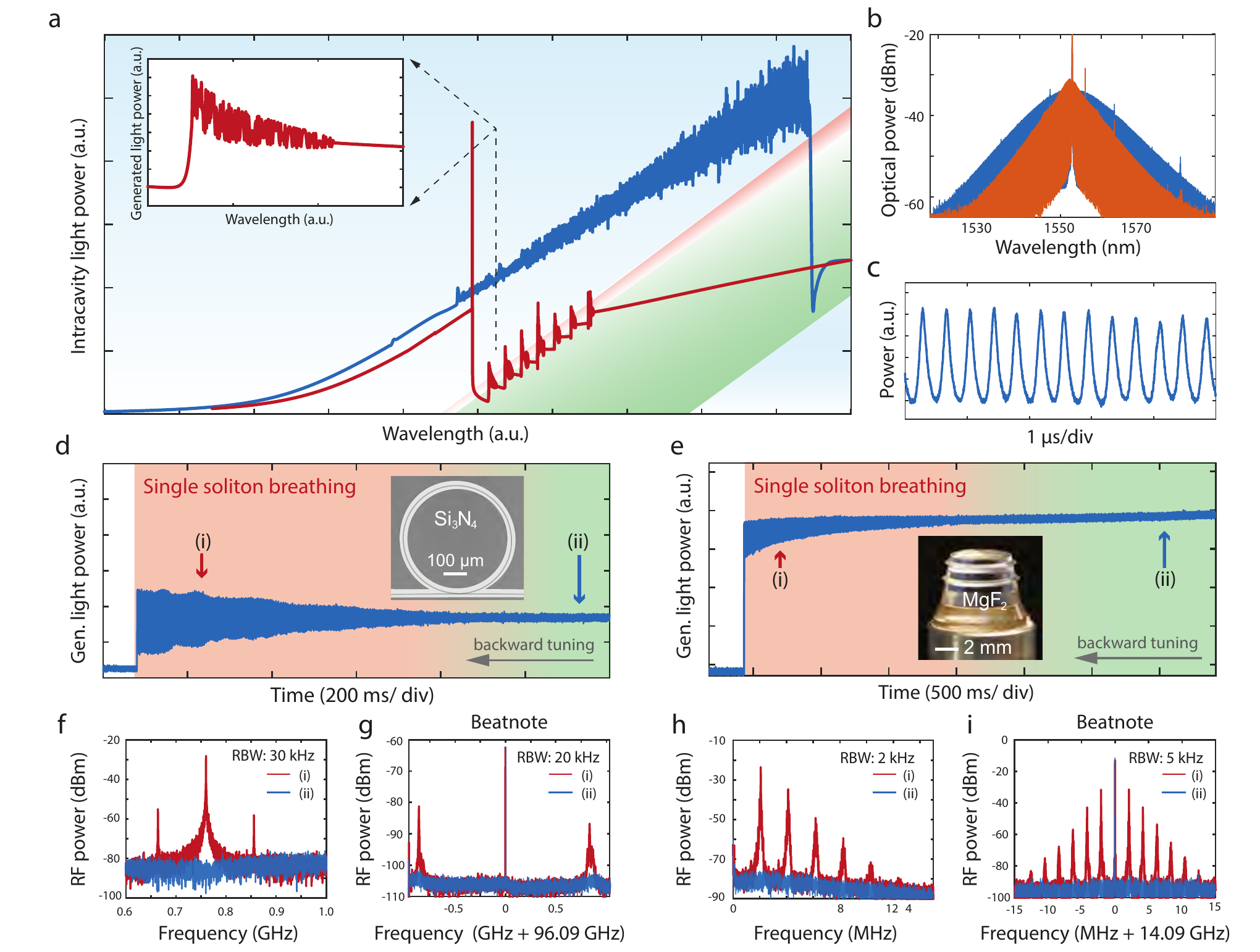}
\protect\caption{\textbf{\boldmath Experimental observation of breathing solitons in $\mathrm{Si_3N_4}$ and  $\mathrm{MgF}_2$ platforms} 
\textbf{(a)} Simulation of the intracavity power, showing the backward tuning method for the excitation of breathing DKS . The generation of a stable multiple soliton state is achieved using forward tuning of the pump laser (blue curve). The backward tuning is applied next (red curve), which enables reaching the low-detuning boundary of the soliton existence range, where the breathing regime (increased noise) and switching effect (step features) occur. The inset details the single soliton breathing and switching. The blue area corresponds to the region where modulation instability occurs, the green marks stationary DKS states and the red area indicates breathing.
\textbf{(b)} Experimental optical spectra of a stationary (blue) and breathing soliton states (red), in the 14~GHz FSR $\mathrm{MgF}_2$ crystalline resonator. The effective detuning $\delta$ is varied by 0.5~MHz between the two states. 
\textbf{(c)} Oscillations of the generated comb power in a breather state of the $\mathrm{MgF_{2}}$ optical resonator resolved with a fast photodiode and high sampling.
\textbf{(d)} Generated-light power evolution, for a single soliton state in the 100~GHz $\mathrm{Si_3N_4}$ microresonator, when the pump is tuned backward, showing the transition from stationary state to breathing and final decay. The inset shows an SEM image of the used microresonator.
\textbf{(e)}  In the $\mathrm{MgF}_2$ crystalline resonator (inset), the comb light evolution features a similar behavior as in (d), when tuning backward.
(\textbf{f}, \textbf{h}) RF spectra of the generated light for a breathing (i, red) and stationary (ii, blue) soliton state respectively in the $\mathrm{Si_3N_4}$ and $\mathrm{MgF}_2$ resonators. In (f), the 0.4~GHz span is centered at 0.8~GHz, close to the fundamental breathing frequency.
(\textbf{g}, \textbf{i}) Repetition rate beatnote for a breathing (i, red) and stationary (ii, blue) soliton state in the $\mathrm{Si_3N_4}$ and $\mathrm{MgF}_2$ resonators.
\label{fig_1}}
\end{figure*}

We suggest to apply the laser ``backward tuning method'' \cite{karpov2016universal} in order to deterministically access the breathing regime in microresonators. Recently, a similar approach was independently employed in fiber cavities \cite{anderson2016chaos}. First, in this approach, a stationary multiple soliton state is excited by sweeping the continuous-wave (CW) driving laser “forward“ (toward longer wavelengths) over the pumped resonance and stopping on the effectively red-detuned side, where solitons are sustained \cite{herr2014soliton}. Second, the driving laser is tuned “backward” (toward shorter wavelength), thus reducing the effective detuning. Due to the microresonator thermal nonlinearity that lifts the fundamental degeneracy of multiple soliton states, this approach was shown to enable the reduction of the intracavity soliton number, via successive switchings to states with a smaller soliton number, thus enabling single soliton access \cite{karpov2016universal}. Figure~\ref{fig_1}a shows a simulation of this excitation scheme in the $\mathrm{Si_3N_4}$ microresonator, including the thermal effects. Forward and backward tuning stages are indicated with blue and red colors correspondingly. The system experiences a series of consecutive switchings, as reflected by the stair-like trace of the intracavity power. We observed that the breathing regime is characterized by oscillations in the intracavity power and occurs in the vicinity of the switching points in each step (see inset in Fig.~\ref{fig_1}a). Importantly, the breathing dynamics can be unambiguously characterized only in the single soliton state, where interactions among different solitons (in a multiple soliton state) are avoided.

\vspace{5 mm}
\noindent \textbf{Experimental identification of breathing.}
We experimentally verified our approach in both platforms, where breathing single soliton were generated using the backward tuning method. Despite significant differences in the resonators properties (\emph{Q} factor, free spectral range (FSR), material dispersion and nonlinearity, cf. Methods), both systems behave qualitatively similarly when approaching and entering the breathing regime.
Figures~1d,e show the experimental evolution of the generated light power of a single soliton in $\mathrm{Si_3N_4}$ and $\mathrm{MgF}_2$ optical resonators when the backward tuning is applied. The signal is obtained by detecting the out-coupled light, after attenuation of the strong pump laser with a narrow fiber Bragg grating notch filter. In both cases the system evolved from a stationary DKS on the right of each trace, to breathing DKS, and finally switched to a homogeneous background, without soliton.
In both platforms, reaching the breathing regime coincides with a progressively increased amplitude noise of the generated light power. 
A detailed measurement (with an increased sampling rate) reveals that the power is oscillating, as shown in Fig.~\ref{fig_1}c.

The oscillatory nature of the out-coupled pulse train in the breathing state can also be characterized by measuring the radio frequency (RF) spectrum. Figures~\ref{fig_1}f,h show the low frequency RF spectra of the stationary and breathing DKS in both optical resonator platforms, at points marked in Fig.~\ref{fig_1}d,e. The stationary soliton state (blue traces) corresponds to a low-noise state of the system, while the breathing state reveals sharp peaks indicating the fundamental breathing frequency and its harmonics (red traces). For our systems, the breathing frequencies were in the range of 0.5--1~GHz for $\mathrm{Si_3N_4}$ microresonators (free spectral range (FSR) $\sim 100~\mathrm{GHz}$) and 1--4~MHz for $\mathrm{MgF}_2$ platform (FSR $\sim 14~\mathrm{GHz}$).

The breather regime can also be evidenced when measuring the repetition rate beatnote. The oscillating pulse dynamics gives rise to additional sidebands around the repetition rate, spaced by the breathing frequency (cf. Figures~1g,i which compares stationary and breathing states in both platforms).

Another characteristic signature of the breathing state is observed in the optical spectrum. Figure \ref{fig_1}b shows the measured spectra of both stationary and breathing single soliton based frequency combs, in a $\mathrm{MgF}_2$ resonator. In the stationary state, the spectrum has a squared hyperbolic secant envelope corresponding to the stationary soliton solution, while in the breathing state, the spectrum features a \emph{triangular} envelope (on a logarithmic scale), resulting from the averaging of the oscillating comb bandwidth, by the optical spectrum analyzer.

\begin{figure*} 
\includegraphics[width = \textwidth]{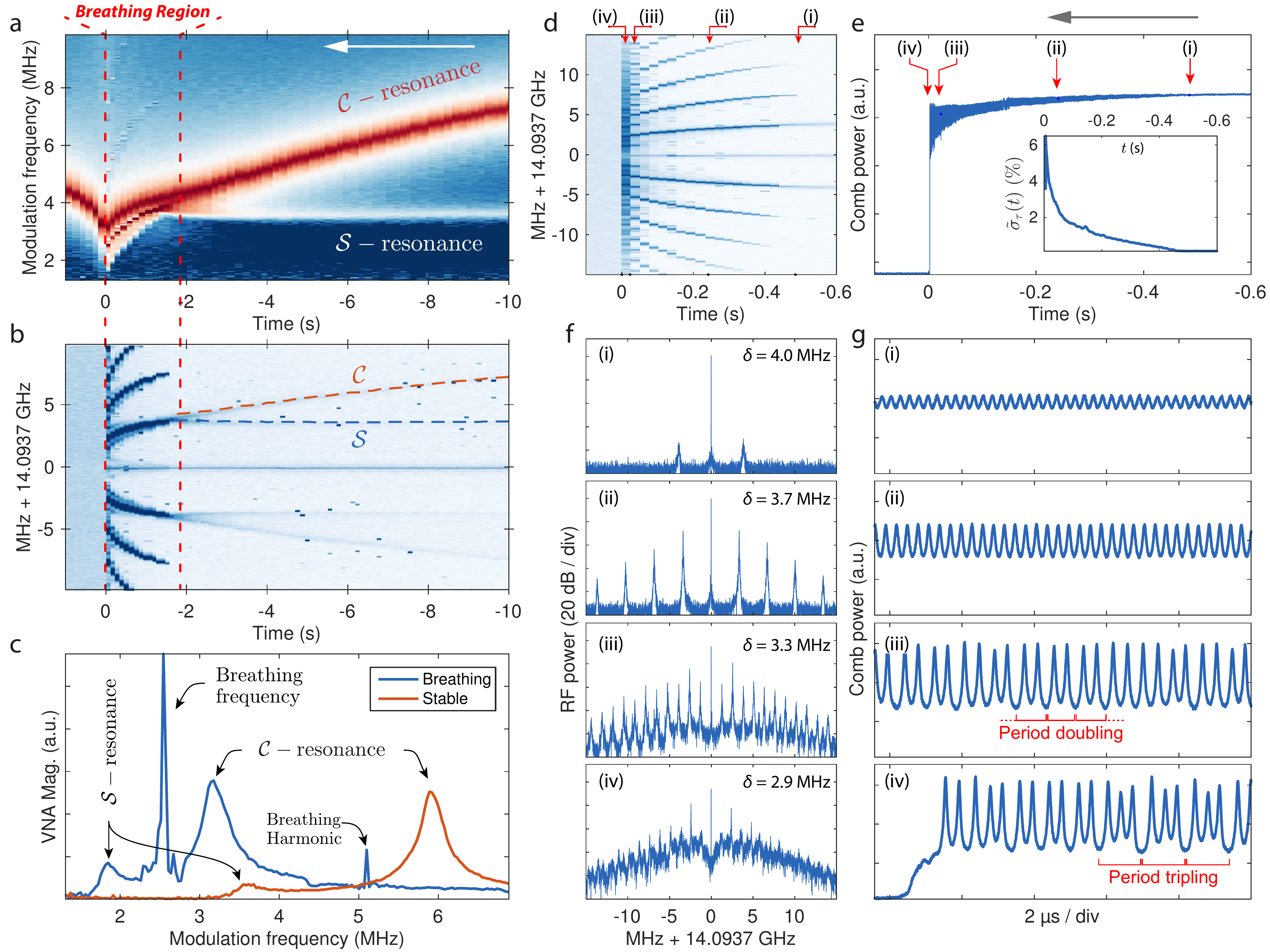}
\protect\caption{\textbf{\boldmath Evolution of the breathing dynamics of a single soliton in backward tuning in the  $\mathrm{MgF}_2$ resonator}
\textbf{(a)} Map of concatenated Vector Network Analyzer (VNA) traces showing the evolution of the modulation response (log scale) from stationary soliton on the right of the time axis to breathing and decay on the left (the time origin $t=0$~s, is set at the soliton decay). As the laser is tuned toward shorter wavelength, the effective detuning ($\mathcal{C}$-resonance) is reduced. The breathing starts typically when the $\mathcal{C}$- and $\mathcal{S}$-resonances separation is on the order of the resonator linewidth.
\textbf{(b)} Corresponding spectrum of the comb repetition rate heterodyne beatnote. 
The modulation response measured on the VNA 
is also visible in the noise of the RF beatnote spectrum
(the dotted lines correspond to the $\mathcal{C}$- and $\mathcal{S}$-frequencies determined on the VNA). The breathing is indicated by the formation of sidebands around the repetition rate beat. As the detuning is reduced, the breathing frequency decreases until the soliton is lost.
\textbf{(c)} Recorded modulation response in the state of a breathing and stationary soliton (linear scale).
\textbf{(d)} Close-in spectrogram of the repetition rate beat within the breathing region and
\textbf{(e)} Corresponding generated comb light evolution. The inset shows the evolution of the relative standard deviation.
\textbf{(f)} Repetition rate beat spectra in the various breathing stages highlighted in (d,e).
\textbf{(g)} Recording of generated light power oscillations at the points highlighted in (d,e).
\label{fig_2}}
\end{figure*}

\begin{figure*}
\includegraphics[width = \textwidth]{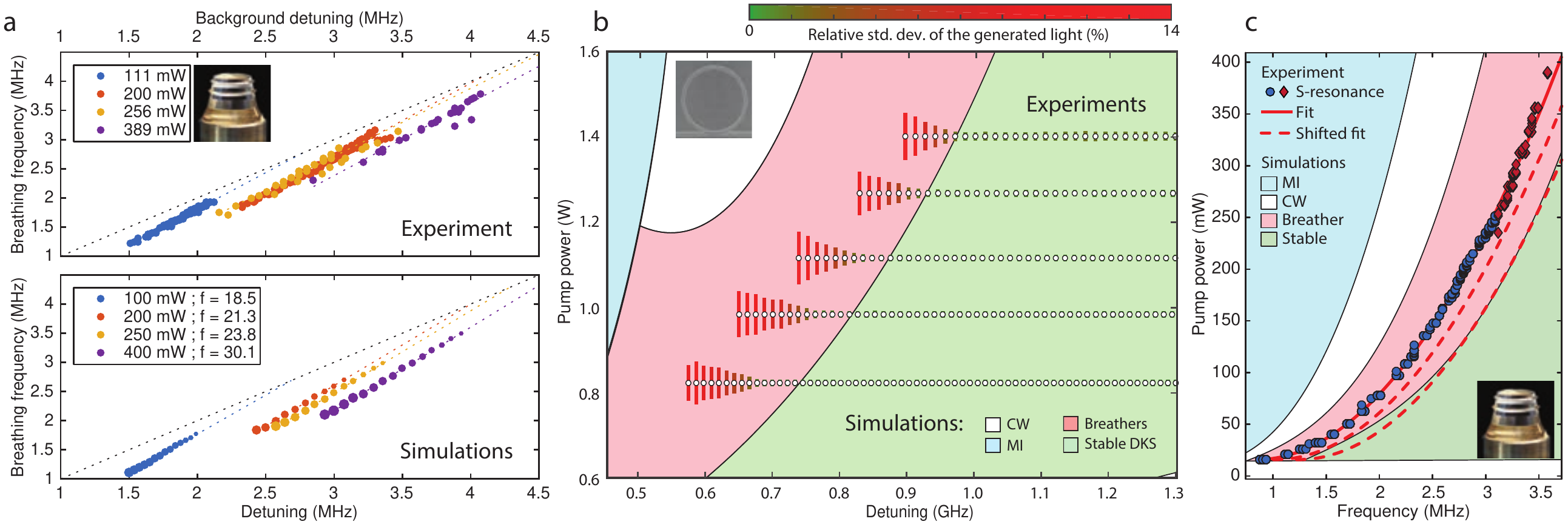}
\protect\caption{\textbf{Breathing dynamics and dependence on the system's parameter}
\textbf{(a)} Top: Experimental determination of the breathing frequency evolution with the detuning for different pump powers, retrieved by the modulation response measurement. Bottom: Simulated evolution of the breathing frequency (see Methods for details).
\textbf{(b)} Stability chart of the CW-pumped $\mathrm{Si_3N_4}$ microresonator in pump power over effective detuning coordinates. White filled circles indicate experimentally accessed DKS states. The colour-coded vertical lines indicate the measured relative standard deviation of the generated light power. The colored background regions and boundaries are interpolated from simulation results (see Methods for details) and correspond to: CW-state (white) -- the soliton decays to the homogeneous background, chaotic modulation instability (blue), stationary DKS state (green) and breathing DKS state (red).  
\textbf{(c)} Dependence of the $\mathcal{S}$-resonance on the pump power, measured for a stationary soliton. This resonance provides an estimate of the detuning point at which the breathing starts. The measurements were carried out with the detuning stabilized to $\delta=3.5$~MHz for $P<300$~mW (blue dots) and $\delta=7$~MHz  $P>300$~mW (red marks). The evolution fits to a parabolic dependence. The regions were retrieved with a polynomial fitting of the boundaries from a simulated stability chart (see Methods for details). The Hopf boundary obtained from the simulations is contained within a  $[+1,+2]\,\kappa/2\pi$ margin (dashed lines) to the measured $\mathcal{S}$ frequency, which is in agreement with the experimental observations reported in Fig.~\ref{fig_2}a.
\label{fig_3}}
\end{figure*}

\begin{figure*}
\includegraphics[width = \textwidth]{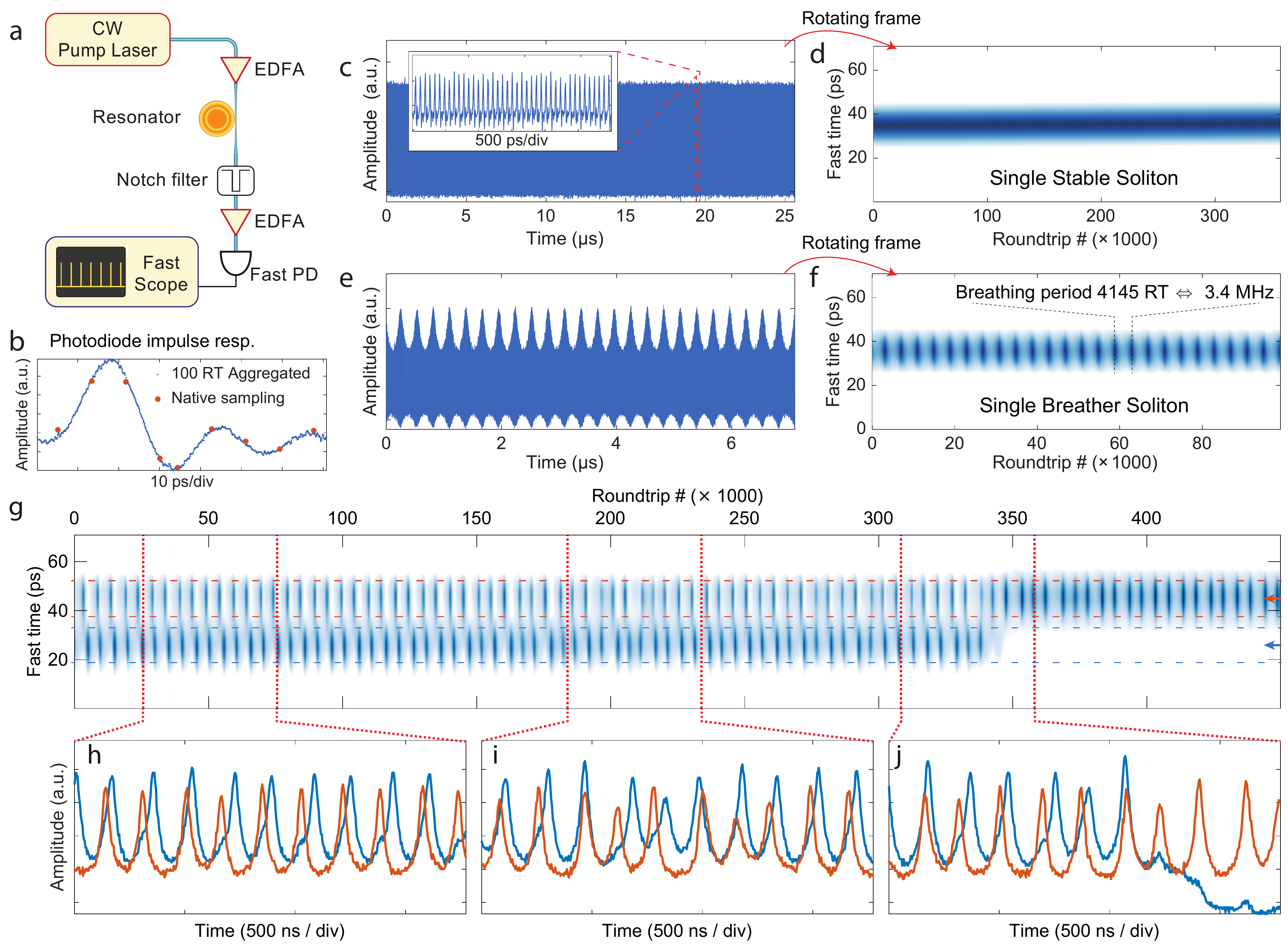}
\protect\caption{\textbf{Direct observation of DKS dynamics}
\textbf{(a)} Experimental setup. Erbium doped fibre amplifier (EDFA) ; Photodiode (PD). 
\textbf{(b)} Photodiode response. The red dots mark the original sampling over a single roundtrip period (RT). With 9 points per period, the pulse amplitude cannot be accurately resolved. This problem is solved by aggregating 100 roundtrips to increase the effective sampling rate and retrieve the impulse response to a single soliton.
\textbf{(c)} Single soliton pulse train, containing $3.5 \times 10^{5}$ roundtrips. The inset shown a short section of the trace, where individual pulses can be coarsely located.
\textbf{(d)}. Dividing the trace in groups of 100 aggregated roundtrips, and stacking reveals the spatiotemporal evolution of the soliton. The soliton position and amplitude is fixed as the soliton is stable. In this map, the colormap is set to remove the ripples of the photodiode response.
\textbf{(e)} Single breathing soliton pulse train.
\textbf{(f)} Applying the same procedure as in (d) reveals the oscillating pulse amplitude while its position remains stable.
\textbf{(g)} Spatiotemporal evolution of a breathing two-solitons state undergoing a transition to a breathing single-soliton state (`switching'). The insets h-j show the evolution of the amplitude of each soliton.
\textbf{(h)} Traces showing a  $\pi/2$ phase difference between the breathing oscillations of the solitons.
\textbf{(i)} Unstable breathing, after which the quadrature relation is restored. 
\textbf{(j)}  Collapse of one soliton, while the other `survives' and remains in the breathing region.
\label{fig_4}}
\end{figure*}

\vspace{5 mm}
\noindent \textbf{Breathing dissipative solitons dynamics.}
Having established a deterministic access to breathers, we next characterized the breathing dynamics. We use a Vector Network Analyzer (VNA) to acquire the system's transfer function from pump phase modulation to the transmitted power, which enables to determine the effective laser detuning of the driven nonlinear system  \cite{karpov2016universal,matsko2015feshbach}.
In the stationary soliton state, this transfer function exhibits two resonances (red curve in Fig.~\ref{fig_2}c) that reflects the bistable nature of the intracavity field (soliton and CW background). The first one ($\mathcal{C}$-resonance) corresponds to the background Kerr-shifted cavity resonance, and indicates the effective pump laser detuning $\delta$ with good approximation. The second one ($\mathcal{S}$-resonance) corresponds to a resonant response of the soliton to the pump modulation. It emerges at lower frequency, and is weakly dependent on the pump laser detuning.

Figure~\ref{fig_2} shows the evolution of a single soliton in a $\mathrm{MgF}_2$ resonator while tuning backward from the stationary state (pump power of 200 mW). During the scan, the system's transfer function is monitored simultaneously with the comb repetition rate beatnote and total comb power.
As the laser detuning is reduced, the $\mathcal{C}$-resonance consequently shifts to lower frequencies (Fig.~\ref{fig_2}a). Interestingly, both $\mathcal{C}$- and $\mathcal{S}$-resonances are also observed in the comb beatnote measurement, appearing as features on the background noise of the electronic spectrum analyzer  (dashed lines in Fig.~\ref{fig_2}b). We ascribe this effect to the transduction of laser input noise via the systems response (i.e. incoherent response which is identical to the probed coherent response). 
The transition from stationary to breathing soliton occurs when the $\mathcal{C}$- and $\mathcal{S}$-resonances separation is on the order of the linewidth ($\kappa/2\pi$), for a detuning $\delta \sim 4$~MHz. Afterward, in the breathing region, strong sidebands at the soliton breathing frequency and its harmonics emerge around the beatnote. The sidebands move progressively closer to the beatnote, revealing that the breathing frequency decreases for smaller detuning. In the breathing state, the transfer function (blue curve in Figure~\ref{fig_2}c) features a strong sharp peak at the breathing frequency that appears in between the $\mathcal{C}$ and  $\mathcal{S}$-resonances. From this response, the breathing frequency and the effective laser detuning can thus be measured with a good precision. Notably, the $\mathcal{S}$-resonance behavior is greatly modified in the breathing domain as it shifts together with the breathing frequency and detuning. We also observed that the transition into the breather regime is reversible by tuning the laser forward (back into the stationary state).

Figures~\ref{fig_2}d--g show the detailed breathing dynamics within the breathing region. 
In particular, the comb power is measured in two ways. First, the global evolution is monitored continuously on a DC coupled photodiode with a slow sampling of $\sim 100$ kSa/s (Fig.~\ref{fig_2}e). Since the breathing oscillations are faster than this sampling rate, they appear as increased amplitude noise, which can be quantified with the relative standard deviation $\tilde{\sigma}_\tau(t) = \sigma_\tau(t) / \mu_\tau(t)$, where $\sigma_\tau$ and $\mu_\tau$ are the local standard deviation and mean power level over $\tau = 1000$ samples.
Second, the fast dynamics resolving the intracavity soliton is also recorded on a real-time oscilloscope with 120 GSa/s, but in short sequences spread over the scan. The breathing pattern in each sequence is then recovered by detecting the envelope of the resolved ``pulse train'' and down sampled (Fig.~\ref{fig_2}g).

The breathing starts with a weak oscillation of the soliton pulse train power (stage i, $\delta \sim 4$~MHz). This corresponds to a single pair of weak sidebands on the comb beatnote. For smaller detuning, the breathing becomes stronger, so that the first sidebands (fundamental breathing frequency) increase, and breathing harmonics emerge (stage ii) as the breathing pattern is not sinusoidal.
At $\delta \sim 3.3$~MHz (stage iii) the system exhibits a \emph{period doubling}, which corresponds to the appearance of sub-sidebands with frequency half of the initial breathing frequency. At last, the breathing turns into strong and irregular oscillations (stage iv, $\delta \sim 2.9$~MHz), exhibiting sporadic transitions to period tripling. This coincides with a large increase in the noise pedestal around the beatnote, although the fundamental breathing frequency remains distinguishable. Finally, the soliton decays quickly thereafter. Such transitions to higher periodicity, temporal chaos, and the collapse match the predicted evolution from numerical studies of the LLE \cite{Leo2013,gelens2014eigenvalue}.

The combined effect of increased modulation depth and reduced breathing frequency is reminiscent of a typical characteristic of complex dynamical systems approaching critical transitions \cite{Scheffer2009,Tredicce2004}. Early warning signals in the form of a variance increase and a critical slowing down have been reported in a wide variety of systems approaching a tipping point, ranging from lasers near threshold to entire ecosystems and the climate \cite{Dakos2008,Carpenter2011,Veraart2012}.

We next studied the breathing frequency as a function of laser detuning at various pump powers, for a single soliton state in the $\mathrm{MgF}_2$ resonator. The backward tuning over the breathing region was repeated for different pump power levels, and the breathing frequency was measured as a function of effective laser detuning (Fig.~\ref{fig_3}a) using the system's transfer function.
The detuning dependence is close to linear: ${f_{b} \approx 1.23 \, \delta + f_b^0}$, where ${f_b}$ is the breathing frequency and $\delta$ indicates the effective laser detuning. The offset $f_b^0$ is observed to decrease with the pump power. We performed numerical simulations based on the LLE, and obtained an almost identical result matching both qualitatively and quantitatively (Fig.~\ref{fig_3}a). A direct linear relation between the breathing frequency and the detuning is also suggested by the approximate breather expression we derived analytically (see Methods).

\vspace{5 mm}
\noindent \textbf{Breathing region.} 
We experimentally studied and mapped the stability chart of DKS solitons in the two-parameter space (pump power $P_{\rm in}$ and effective detuning $\delta$) of the CW-pumped microresonator system \cite{godey2014stabilityLLE, leo2010temporal, Jaramillo2015deterministicSoliton}.
A stationary single soliton was generated using the backward tuning method at different pump powers, and gradually tuned across the breathing region until its decay. The white circles in Fig.~\ref{fig_3}b mark the operating points ($P_{\rm in}$, $\delta$) thus accessed experimentally. The color-coded vertical line around each circle indicates the relative standard deviation of the output power measured at the corresponding point and directly relates to the breathing amplitude. The results reveal a pump power dependency of the breathing region, which location shifts towards higher effective detuning values and range slightly reduces, as the pump power increases.

We compared our experimental results to LLE-based simulations (see Methods for details). The resulting types of intracavity field attractors at various operating point are labeled via the background color-coding in Fig.~\ref{fig_3}b: CW-state (white color), where the soliton decays to the homogeneous background; chaotic modulation instability state (blue); stationary soliton state (green); and breathing soliton state (red). The experimentally accessed stationary and breathing states are well within the corresponding areas predicted by the simulations. The mismatch between experimental results and simulations for the low-detuning boundary can be attributed to the deviations between the measured detuning values and the true $\delta$ that differ at low-detuning due to the higher background \cite{karpov2016universal}. The highly unstable and short lived breather in this region makes it harder to resolve. Finally, high-order dispersion and nonlinear effects (e.g. Raman scattering, avoided mode crossings and third order dispersion) were not-included for simplicity in the simulations, but are present in the real microresonator system.

Furthermore, as noted earlier, the breathing emerges when the $\mathcal{C}$-resonance is tuned close to the $\mathcal{S}$-resonance, and their separation is on the order of the resonator linewidth. Therefore the $\mathcal{S}$-resonance frequency provides an estimate for the detuning value of the upper boundary of the breathing region (Hopf bifurcation). Experimentally, we monitor the $\mathcal{S}$-resonance frequency as the pump power is raised, while stabilizing the laser detuning ($\mathcal{C}$-resonance frequency) to a constant value in the stationary soliton state \cite{lucas2016study}. Figure~\ref{fig_3}c reports the evolution of the $\mathcal{S}$-resonance frequency with the pump power for the $\rm MgF_2$ resonator, whose smaller linewidth produces narrower resonance peaks in the transfer function that are easily resolved.
The obtained relation fits to a parabolic dependence and matches the Hopf boundary retrieved from simulations with a frequency offset that does not exceed twice the linewidth, showing that the breathing region can be identified even from the stationary state.

\vspace{5 mm}
\noindent \textbf{Real-time observation of breathers.}
The soliton dynamics in the microresonator is studied further in the time domain by measuring the 14~GHz soliton pulse train coupled out of the $\rm MgF_2$ resonator. The generated light is amplified and detected on a fast photodiode (70~GHz bandwidth) connected to a real-time oscilloscope with 45~GHz analog bandwidth (sampling rate 120 GSa/s). We note that so far, the real time sampling of successive solitons in microresonators had not been attained due to the required high sampling bandwidth. The present configuration allows for the measurement of $\sim 9$ samples per roundtrip and enable a coarse localization of the soliton pulse within one roundtrip as shown in Figure~\ref{fig_4}b,c.
Since we observe that the soliton breathing dynamics evolves over a large number of roundtrips ($>1000$), we aggregate together the samples contained in segments of 100 roundtrips, to achieve an effectively larger sampling rate. This produces smoother traces, revealing the impulse response of the acquisition system (matching with the photodiode response), where especially the instantaneous soliton peak amplitude can be reliably retrieved (Fig.~\ref{fig_4}b). Longer traces (Fig.~\ref{fig_4}c,e) that measure the evolution over a large number of roundtrips are divided in 100-roundtrips segments, aggregated and stacked. This facilitates the visualization of a spatiotemporal evolution of the soliton amplitude and intracavity position in the rotating frame.

We first benchmarked our measurement procedure in the single soliton state. At a pump power of 230 mW and for the effective laser detuning $\sim10$~MHz, the soliton is stationary as expected, with a constant amplitude (Fig.~\ref{fig_4}d). For a smaller detuning $\sim3.5$~MHz, the breathing soliton is generated, and the time trace reveals the oscillatory envelope of the soliton amplitude (Fig.~\ref{fig_4}e). In the rotating frame, this leaves a dotted pattern at the breathing period (Fig.~\ref{fig_4}f), where the breathing frequency is $\sim3.4$~MHz corresponding to 4145 roundtrips.

The fast recording on the real-time oscilloscope also enables to delineate the breathing dynamics of individual pulses in a multiple soliton state. Figure~\ref{fig_4}g shows the evolution of a breathing two-solitons state, during a backward tuning around $\delta\sim2.1$~MHz. The state experiences a switching \cite{karpov2016universal}, where one soliton decays and the other survives.  Furthermore, in this small detuning condition, the breathing is typically irregular and might be locally identified as period doubling or tripling, as reflected on the traces (Fig.~\ref{fig_4}g--j). The measurement reveals that the two solitons breathe overall at the same frequency, but are not in phase. In the present case, there seems to exist a preferred phase relation of $\sim \pi/2$. Even if the breathing is irregular and the phase relation can be locally altered as shown in Fig.~\ref{fig_4}i, the relative phases seem to quickly recover this relation. Such behavior has been predicted by Turaev et al. \cite{Turaev2012}, showing that the longer interaction length of breathing soliton can lead them to form bound states with a specific inter-distance and breathing phase relation. A quadrature breathing should correspond to a comparatively large soliton separation, which matches with the above case as the pulse are separated by more than the photodiode response time.

\section{Discussions}
We have experimentally demonstrated the formation of breathing dissipative solitons in two distinct microresonator platforms -- ${\rm MgF_2}$ crystalline resonator and photonic chip ${\rm Si_3N_4}$ microresonator -- having different characteristics, which validates the universal nature of our observations. We demonstrated a laser tuning method which enables a reliable access to soliton breathing. Typical signatures of breathing solitons, including a periodic varying soliton peak intensity and a triangular spectral envelope are observed. % Especially, the method allows for accessing the single breathing soliton state
Moreover, we presented  a direct time-resolved observation of dissipative Kerr solitons in microresonators, revealing the breathing dynamics of individual solitons in both single and multiple breathing soliton states. Such measurements unambiguously reveal the transition to higher breathing periodicity and more chaotic type of behavior.
By monitoring the laser detuning of the driven nonlinear system,  we were able for the first time, to experimentally measure the breathing frequency and its dependence on the laser detuning. These studies evidenced a linear relation, which agrees remarkably well with numerical simulations, and provides further insights into this breathing parameter.
Furthermore, the experimental mapping of the transition boundary from stationary state to breathing state, reveals a parabolic-like relation between the pump power and the detuning, which also matches numerical simulations.
In the context of low-noise operation of soliton-based microresonator frequency combs, breathing degrades the soliton stability, and should be generally avoided. Our results provide useful diagnosis tools to determine the breathing boundary, even from the stationary soliton state.
These findings not only carry importance from an application perspective, but also contribute more broadly to the fundamental understanding of dissipative soliton physics.
Our observations further highlight the suitability of the microresonator platform for the study of nonlinear dynamics, especially for accessing high normalized driving values. In the present case, the remarkable agreement between the numerical simulations and experimental observations validates the relevance of the numerical models, even in such cases of non-stationary and chaotic dynamics.

\section*{Methods}
\label{Methods} 
\noindent\textbf{Optical resonators}: $\mathcal{\mathcal{\mathrm{Si_{3}N_{4}}}}$
integrated microring resonators with the free spectral range (FSR) of $\sim100$ $\mathrm{GHz}$ and Q-factors $\sim10^6$ (linewidth $\frac{\kappa}{2\pi}=150-200~{\rm~MHz}$)  was fabricated using the
Photonic Damascene process \cite{Pfeiffer2015Damascene}. In order to achieve the single mode operation and suppress the effect of avoided mode crossings, a ``filtering section'' was added to the microresonator \cite{Herr2014avoidmodecross,kordts2015higher}. The resonators dispersion parameters -- ${\frac{D_{2}}{2\pi}=2~{\rm~MHz}}$, ${\frac{D_{3}}{2\pi}=\mathcal{O}(1~{\rm kHz})}$ -- were measured using the frequency comb assisted laser spectroscopy method \cite{Delhaye2009disp} (the resonance frequencies near ${\omega_{0}}$ are expressed
in a series ${\omega_{\mu}=\omega_{0}+\sum_{i\ge1}D_{i}\mu^{i}/{i!}}$,
where ${i\in\mathbb{N}}$, ${\mu\in{\mathbb{Z}}}$ is the mode number). 
The wavelength of CW pump laser in experiments was set at $1553~{\rm nm}$, the pump power varied from $1$ to $4~{\rm W}$.

The $\mathrm{MgF_{2}}$ crystalline resonator with FSR $\frac{D_{1}}{2\pi}=14.094$~GHz was fabricated by diamond turning of a cylindrical blank. The high Q-factor of $\sim10^9$ (intrinsic linewidth $\frac{\kappa_0}{2\pi}=80~{\rm kHz}$, intrinsic finesse $\mathcal{F} \sim 1.7 \times 10^{5}$) was achieved with subsequent hand polishing.
The dispersion parameters at the pump wavelength of $1553$ $\mathrm{nm}$ are: ${\frac{D_{2}}{2\pi}=1.96~{\rm kHz}}$,
${\frac{D_{3}}{2\pi}=\mathcal{O}(1~{\rm Hz})}$.
The pump laser (fiber laser, wavelength $1553$ $\mathrm{nm}$; short-term linewidth $10$~kHz) is amplified between $\sim20 \text{ and } 450~{\rm mW}$ and evanescently coupled to the resonator with a tapered optical fiber, which enables a tuning of the coupling. The loaded linewidth $\kappa$ is retrieved by measuring the $\mathcal{C}$-resonance linewidth in the VNA trace (when no solitons are present in the cavity), the associated coupling coefficient $\eta = \kappa_{\rm ex}/\kappa = (\kappa - \kappa_{0})/\kappa$ was measured in the range 0.45 to 0.62.

\vspace{5 mm}
\noindent\textbf{Numerical simulations:}
Numerical simulation based on the LLE were implemented in order to study breathers. For the $\rm MgF_2$ resonator, the simulations are performed using periodic boundary conditions with 1024 discretization points (1024 modes). The simulation of the breathing frequency as a function of the control parameters is carried as follow: The operating parameters (pump power and the laser detuning) are fixed and the intracavity field is initiated with a single (stationary) soliton ansatz \cite{herr2014soliton}. The simulation of the soliton evolution is then carried over 15 photon lifetimes ($2\pi/\kappa$) and the breathing dynamics analysis is carried over the final 2/3 time range, where the stationary soliton ansatz is converged to system's inherent breathing state. The oscillation frequency is determined via spectral analysis and plotted in Fig.~\ref{fig_3}a.
The simulations of the stability chart of the $\mathrm{Si_3N_4}$ microresonator presented in Fig.~\ref{fig_3}b were performed with 512 modes. Using hard excitation scheme, stationary DKS were seeded at fixed input powers and large detunings.  Then the laser detuning was reduced step by step to map over the chart. In each step, the intracavity field pattern is characterized after $\sim5000$-roundtrips to exclude early-stage transient formations.
In simulations for both $\rm MgF_2$ and $\mathrm{Si_3N_4}$ microresonators, we identified intracavity states of single stationary soliton state, breathing soliton state, chaotic state in the operation regime of modulation instability (MI) and state where intracavity field decays leaving only the cw background, showing in color-codings in Fig.~\ref{fig_3}b,c.

\vspace{5 mm}
\noindent\textbf{Approximate breather ansatz:}
We develop an approximate breather solution for the LLE (1), that allows to inspect the relation of the breathing regime parameters to the pump power and the effective detuning.
It is known that an approximate stationary solution of the LLE for positive $\zeta_0$ (i.e. for the pump laser being effectively red detuned) may be found as a sum of the soliton and a background:
\begin{eqnarray}
\Psi(\theta) \approx \Psi_{C}+\Psi_{\cal S}(\theta) e^{i\phi_0}, 
\end{eqnarray}
here  $\Psi(\theta)$ is the intracavity waveform,  $\theta=\phi\sqrt{\frac{1}{2d_{2}}}$ is the dimensionless longitudinal coordinate, $\phi$ is the co-rotating angular coordinate of the resonator and  $d_{2}=D_{2}/\kappa$ is the dimensionless dispersion.
$\Psi_{\cal C}\approx -i f/\zeta_0$ represents the constant solution of \eqref{eq:nls} (background), while $\Psi_{\cal S}=B\,\sech(B\theta)$ is the exact stationary conservative soliton solution of \eqref{eq:nls} (without loss nor drive), with $B=\sqrt{2\zeta_0}$. The phase $\phi_0$  may be found by perturbation methods \cite{GreluChapter} from $\cos\phi_0=2B/\pi f$.

The exact Kuznetsov-Ma breather \cite{Kuznetsov1977, Ma1979} solution of \eqref{eq:nls} without loss and pump can be employed to derive an approximate ansatz for dissipative breathing solitons:

\begin{align}
\Psi_{\cal S}(\theta,\tau)&=\left(\frac{K_1\cos\Omega \tau+iK_2 \sin \Omega\tau}{\cosh B\theta-K_3\cos \Omega \tau}-\epsilon\right)e^{iK_4\tau}\nonumber\\
\Omega &= \frac{B}{2}\sqrt{B^2+4\epsilon^2}\nonumber\\
K_1&=\frac{B^2}{\sqrt{B^2+4\epsilon^2}}\nonumber\\
K_2&=B\nonumber\\
K_3&=\frac{2\epsilon}{\sqrt{B^2+4\epsilon^2}}\nonumber\\
K_4&=\epsilon^2-\zeta_0
\end{align}

If the time dependent part of the background $\epsilon$  is small then leaving only terms up to the first order on $\epsilon \to 0$ we arrive at:
\begin{align}
\Psi_{\cal S}(\theta,\tau)&=B\sech(B \theta)+2\epsilon \cos(\zeta_0 t)\sech^2(B\theta)
-\epsilon e^{-i\zeta_0 t}\label{decomb}
\end{align}
We notice, that for $\epsilon=0$ this breather converges to a simple stationary soliton, and for small $\epsilon$ the oscillation frequency of both the background and soliton itself simply coincides with the laser detuning.

\bibliography{SolitonSwitchingLibrary}

\begin{acknowledgments}
\label{Acknowledgements}
The authors gratefully acknowledge V. Brasch, for his help in setting up the fast detection scheme, as well as the valuable discussions and support of J.D. Jost, M.H.P. Pfeiffer, M. Anderson and M. Geiselmann.
This publication was supported by the Swiss National Science Foundation (SNF) under grant agreement 161573
as well as Contract W31P4Q-14-C-0050 (PULSE) from the Defense Advanced Research Projects Agency (DARPA), Defense Sciences Office (DSO), and by the European Space Agency (ESA), European Space Research and Technology Centre (ESTEC), Contract No. 4000116145/16/NL/MH/GM.
This material is based upon work supported by the Air Force Office of Scientific Research, Air Force Material Command, USAF under Award No. FA9550-15-1-0099. 
M.K. acknowledges funding support from EU FP7 programme under Marie Sklodowska-Curie ITN grant agreement No. 607493.
H.G. acknowledges funding support from EU Horizon 2020 research and innovation programme under Marie Sklodowska-Curie grant agreement No 709249.
All samples were fabricated and grown in the Center of MicroNanoTechnology (CMi) at EPFL.
\end{acknowledgments}

\end{document}